\documentclass[twocolumn,pre,showpacs,superscriptaddress]{revtex4} 
\usepackage{bm}
\usepackage{graphicx}%
\usepackage{color} 
\usepackage{transparent}
\usepackage{psfrag}
\usepackage{dcolumn}
\usepackage{amsmath}
\usepackage{amssymb}
\usepackage{latexsym}
\usepackage{hyperref} 
\usepackage{booktabs}
\usepackage{latexsym}
\usepackage{comment}
\begin{document}

\def\bea{\begin{eqnarray}}
\def\eea{\end{eqnarray}}
\def\beq{\begin{equation}}
\def\eeq{\end{equation}}
\def\f{\frac}
\def\k{\kappa}
\def\e{\epsilon}
\def\ve{\varepsilon}
\def\be{\beta}
\def\D{\Delta}
\def\h{\theta}
\def\t{\tau}
\def\a{\alpha}
\def\cDa{{\cal D}[X]}
\def\cD{{\cal D}[x]}
\def\cL{{\cal L}}
\def\cLo{{\cal L}_0}
\def\cLa{{\cal L}_1}
\def\rv{{\bf r}}
\def\tv{\hat t}
\def\on{{\omega_{\rm on}}}
\def\off{{\omega_{\rm off}}}
\def\fv{{\bf{f}}}
\def\fm{\bf{f}_m}
\def\zh{\hat{z}}
\def\yh{\hat{y}}
\def\xh{\hat{x}}
\def\km{k_{m}}
\def\Re{{\rm Re}}
\def\sj{\sum_{j=1}^2}
\def\rk{\rho^{ (k) }}
\def\rek{\rho^{ (1) }}
\def\cek{C^{ (1) }}
\def\rz{\rho^{ (0) }}
\def\rt{\rho^{ (2) }}
\def\rtb{\bar \rho^{ (2) }}
\def\trk{\tilde\rho^{ (k) }}
\def\trek{\tilde\rho^{ (1) }}
\def\trz{\tilde\rho^{ (0) }}
\def\trt{\tilde\rho^{ (2) }}
\def\r{\rho}
\def\tD{\tilde {D}}
\def\s{\sigma}
\def\kb{k_B}
\def\bF{\bar{\cal F}}
\def\F{{\cal F}}
\def\la{\langle}
\def\ra{\rangle}
\def\nn{\nonumber}
\def\up{\uparrow}
\def\dn{\downarrow}
\def\S{\Sigma}
\def\dg{\dagger}
\def\d{\delta}
\def\p{\partial}
\def\l{\lambda}
\def\L{\Lambda}
\def\G{\Gamma}
\def\o{\Omega}
\def\w{\omega}
\def\g{\gamma}
\def\E{{\mathcal E}}
\def\O{\Omega}
\def\vv{ {\bf v}}
\def\jv{ {\bf j}}
\def\jr{ {\bf j}_r}
\def\jd{ {\bf j}_d}
\def\jdd{ { j}_d}
\def\noi{\noindent}
\def\a{\alpha}
\def\d{\delta}
\def\p{\partial} 
\def\la{\langle}
\def\ra{\rangle}
\def\e{\epsilon}
\def\n{\eta}
\def\g{\gamma} 
\def\hf{\frac{1}{2}}
\def\rcurs{r_{ij}}
\def\bv{ {\bf b}}
\def\uv{ {\bf u}}
\def\rv{ {\bf r}}
\def\cf{{\mathcal F}}

\def\dact{\tilde{D}_v}

\title{Extension and dynamical phases in random walkers depositing and following chemical trails}

\author{Subhashree Subhrasmita Khuntia}%
\email{subhashree@iisermohali.ac.in}
\affiliation{Indian Institute of Science Education and Research Mohali, Knowledge City, Sector 81, SAS Nagar 140306, Punjab, India}
\author{Abhishek Chaudhuri}
\email{abhishek@iisermohali.ac.in}
\affiliation{Indian Institute of Science Education and Research Mohali, Knowledge City, Sector 81, SAS Nagar 140306, Punjab, India}
\author{Debasish Chaudhuri}%
\email{debc@iopb.res.in}
\affiliation{Institute of Physics, Sachivalaya Marg, Bhubaneswar 751005, India}
\affiliation{Homi Bhaba National Institute, Anushaktinagar, Mumbai 400094, India}

%

\date{\today}

\begin{abstract}
	{ Active walker models have proved to be extremely effective in understanding the evolution of a large class of systems in biology like ant trail formation and pedestrian trails. We propose a simple model of a random walker which modifies its local environment that in turn influences the motion of the walker at a {\em later} time. We perform direct numerical simulations of the walker in a discrete lattice with the walker actively depositing a chemical which attracts the walker trajectory and also evaporates in time. We propose a method to look at the structural transitions of the trajectory using radius of gyration for finite time walks. The extension over a definite time-window shows a non-monotonic change with the deposition rate characteristic of a coil-globule transition. At certain regions of the parameter space of the chemical deposition and evaporation rates, the extensions of the walker shows a re-entrant behavior. The dynamics, characterised by the mean-squared displacement, shows deviation from diffusive scaling at intermediate time scales, returning to diffusive behavior asymptotically. A mean field theory captures the variation of the asymptotic diffusivity.}
\end{abstract}

\maketitle

\section{Introduction}
The emergence of large scale complexity as a result of interaction between individuals that can only perceive and influence its local neighborhood is one of the most fascinating topics of  biology \cite{Moussaid2009,Olsen2019,camazine2020}. The examples of such phenomena span system sizes from micro to macro scales, including quorum sensing, auto-chemotaxis and biofilm formation in bacteria, formation of neural networks and axon bundles, and formation of ant trails and pedestrian tracks~\cite{Mukherjee2019,John2009,Chaudhuri2015,Helbing1997,Vicsek1994nature,Chaudhuri2011,Robles2018,Feinerman2017,Khuong2016,Gloag2016,king2009chemotaxis,mittal2003motility,jin2017chemotaxis}.
The common factor in all such emergent collective properties is that individual motile agents modify the local environment which in turn influences the dynamics of the agents over time. In bacterial quorum sensing, cells communicate by production and response to extracellular signalling molecules called autoinducers. A fluctuation towards a higher density of cells lead to a build up of higher concentration of autoinducers. This generates a positive feedback accumulating more cells, further increasing the bacterial concentration. Bacteria can use this to alter their gene expression to change their virulence~\cite{Rutherford2012,Laganenka2016}. On the other hand, ants secrete pheromone in their local environment that other ants can sense to decide their direction of motion leading to ant-trail formation. A similar phenomenon is observed in the emergence of pedestrian track. While walking, a pedestrian deforms the vegetation on a field. The deformation remains long after the pedestrian is gone. Other pedestrians follow this deformation while choosing a path on the field, as a result reinforcing the deformation, so that finally a pedestrian track emerges~\cite{Helbing1997,helbing1997modelling}.  

Many of these biological entities, including bacteria and ants, perform active motion consuming energy. This aspect of their non-equilibrium dynamics can be described in terms of simple models like active Brownian particle (ABP) and run and tumble particle (RTP)\cite{Bechinger2016, Marchetti2013, Malakar2018}. The active particles show several counter-intuitive properties including multiple crossovers between ballistic and diffusive motion, accumulation near confining boundaries, etc. \cite{Dolai2020, Santra2021, Shee2020, Chaudhuri2020, Shee2021b, Shee2022a}. The other important aspect of their dynamics is their {\em time-delayed} but spatially local interaction achieved via signalling through autoinducers or pheromone. The process of such signaling is inherently non-equilibrium, associated with secretion of the chemicals and their diffusion and evaporation. The impact of this non-equilibrium delayed interaction on the motion of a walker  is the focus of the current paper. 

Among agent-based models, an active random walker (ARW) model serves as a flexible tool to demonstrate the evolution of the class of systems described above\cite{Lam1995,Castillo1995,Schweitzer2007}. 
In an ARW model, the walkers modify their environment which, in turn, influences them at a later time. In these models, typically, the walker is assumed to perform a directed random motion interacting with the environmental modification it produces. In another approach trajectories of walkers are considered in an fluctuating environment. These approaches have been used to analyze varied problems like the formation of river basins~\cite{rodriguez2001fractal}, and fasciculation of axons in developing neurons~\cite{Chaudhuri2009, Chaudhuri2011e}. Here, instead, we consider a simple random walker to delineate the impact of delayed interactions.  

 Several theoretical models have looked at positive auto-chemotaxis~\cite{Taktikos2011,Kranz2016,Kranz2019,Grima2005,Grima2006,Sengupta2009,Tsori2004,saha2014clusters,marsden2014chemotactic,hokmabad2022chemotactic}, where a particle moves towards regions of higher gradient of chemical secreted by itself. In some approaches, the particles are modelled as persistent walkers moving at a constant speed which is independent of chemical gradient or concentration~\cite{Taktikos2011,Kranz2016,Kranz2019}. It was shown that reduction of persistence in the heading direction due to chemical coupling leads to a reduction of effective active diffusion. In the absence of evaporation, increased coupling with the chemical field could lead to localization with vanishing diffusion coefficient at long times. This is different from approaches where the concentration gradient leads to a mean speed of a random walker where the motion is shown to be diffusive in the long time limit~\cite{Grima2005,Grima2006,Sengupta2009}. Further, it was shown before, in the absence of evaporation, a auto-chemotactic random walker could also get localized~\cite{Tsori2004}.

In our model, we consider a random walker which actively deposits a chemical that is assumed to evaporate over time. This chemical, in turn, sets an attractive potential for the walker to follow in future steps, as in ARW model. The deposition rate $\alpha$ and evaporation rate $\beta$ controls the emergent structure of the trail. We observe that even a single walker displays intriguing structural and dynamical features. The extension of the trails show non-monotonic variation with increasing deposition rate. The initial reduction of extension is reminiscent of the coil-globule transition in polymers. We obtain a detailed phase diagram in the $\alpha-\beta$ plane showing regions of re-entrant behavior for the extension of the walker trajectory and chemical trails. This is the first main result of our paper.
For large enough deposition rates, the mean-squared displacement (MSD) of the random walker shows clear deviation from diffusive scaling over intermediate time scales, returning to diffusive behavior asymptotically, albeit with a decreased diffusion constant. We develop a mean-field description of the motion leading to an estimate of the effective diffusion constant that captures the numerically obtained behavior. This is our second main result.    
 
\begin{figure}[!t]  
 \centering
  \includegraphics[width=0.95\linewidth]{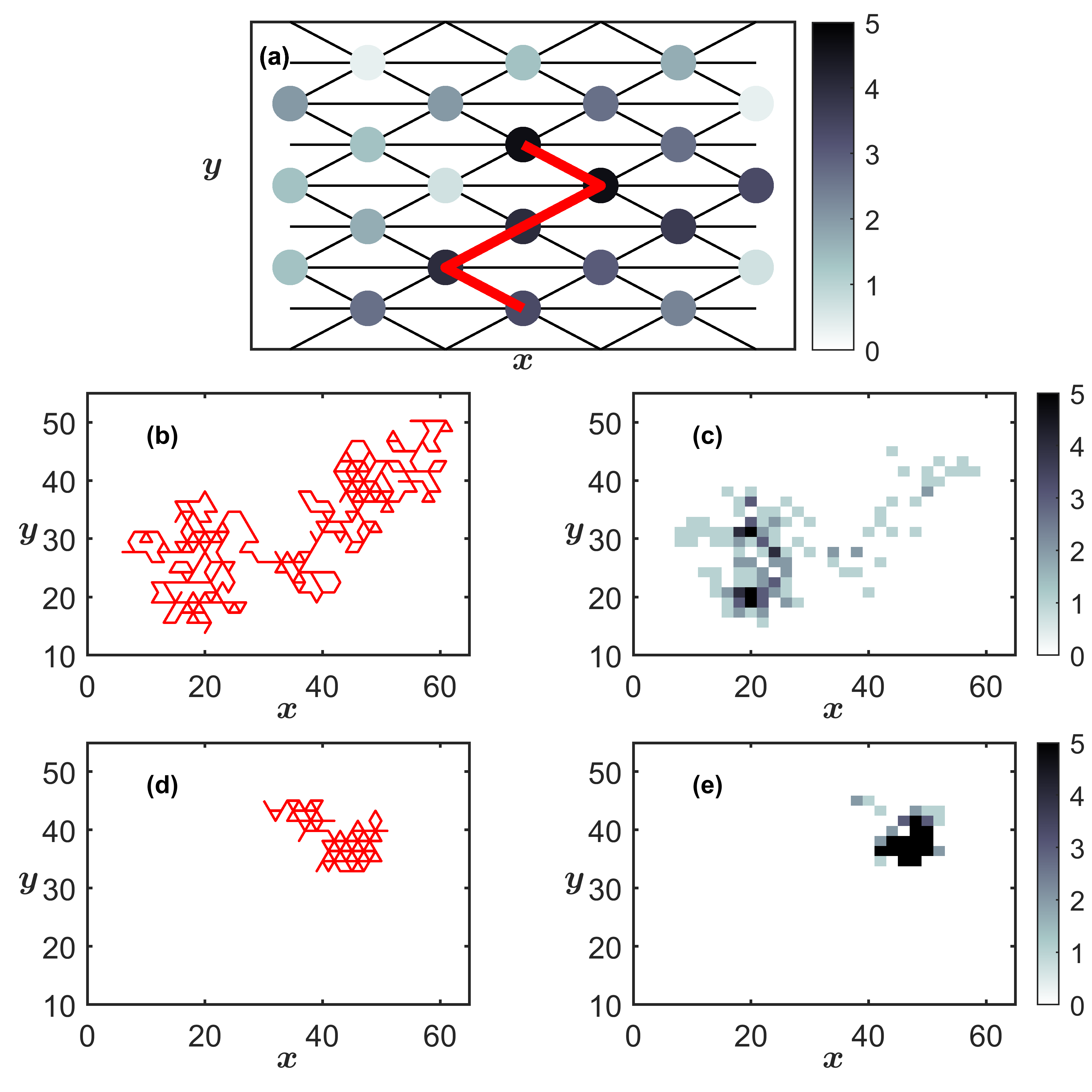}
     \caption{($a$)~Description of the model. The red line denotes a possible trajectory of the random walker on a triangular lattice. The walker can move along any of the bonds. The shades at different nodes denote the strength of deposited chemical. The simulated trajectories of the walker ($b,d$) and configurations of the corresponding chemical trail ($c,e$) obtained at $\alpha=0.0005,0.5$ respectively and $\beta=0.007$.
     }
    \label{fig:config}
  \end{figure}

\section{Model}
We consider a walker performing random walk in two dimensions on a triangular lattice with open boundary as shown in Fig.~\ref{fig:config}($a$). The dynamics of the walker is specified according to the following rules : (1) The walker performs a biased random walk on a two dimensional triangular lattice. (2) The walk gets biased towards particular neighboring sites due to a chemical cue present at these sites. (3) The chemical cue is in turn generated by the walker which deposits a chemical at its current position at a fixed rate $\alpha$. Its concentration $c_i(t)$ evaporates with time with a decay rate $\beta$. The evolution can be expressed as 
    \begin{equation}
        \frac{dc_i(t)}{dt} = \alpha - \beta c_i(t).
    \end{equation}
    Thus the walker can change the state of its local environment. (4) The displacement to one of the six nearest neighbor sites proceeds stochastically with probability   
    \begin{equation}
        p_i(t) = 
        {\cal Z}^{-1}(1+ \nu_i(t)),
        \label{eqn2}
    \end{equation}
    where the bias $\nu_i= \lambda c_i/(1+c_i)$ is parametrized through $\lambda$ and the normalization constant ${\cal Z}=\sum\limits_{j=1}^{nn} (1+ \nu_j(t))$ 
    with $\textrm{nn} = 6$ denoting the number of nearest neighbors. Motivated by the  typical biological response, here the chemical bias is assumed to saturate with increasing concentration $c_i$. The choice of $p_i$ ensures that in the limits of $c_i=0$ or $\infty$ the walker performs unbiased diffusion. This choice is different from, e.g., that in Ref.~\cite{Taktikos2011, Kranz2016, Kranz2019}, where the coupling is assumed to be constant. As we show later, such a biologically motivated choice leads to a re-entrant behavior in the extension of walker trajectory.

  \begin{figure}[!t]
\includegraphics[width=0.95\linewidth]{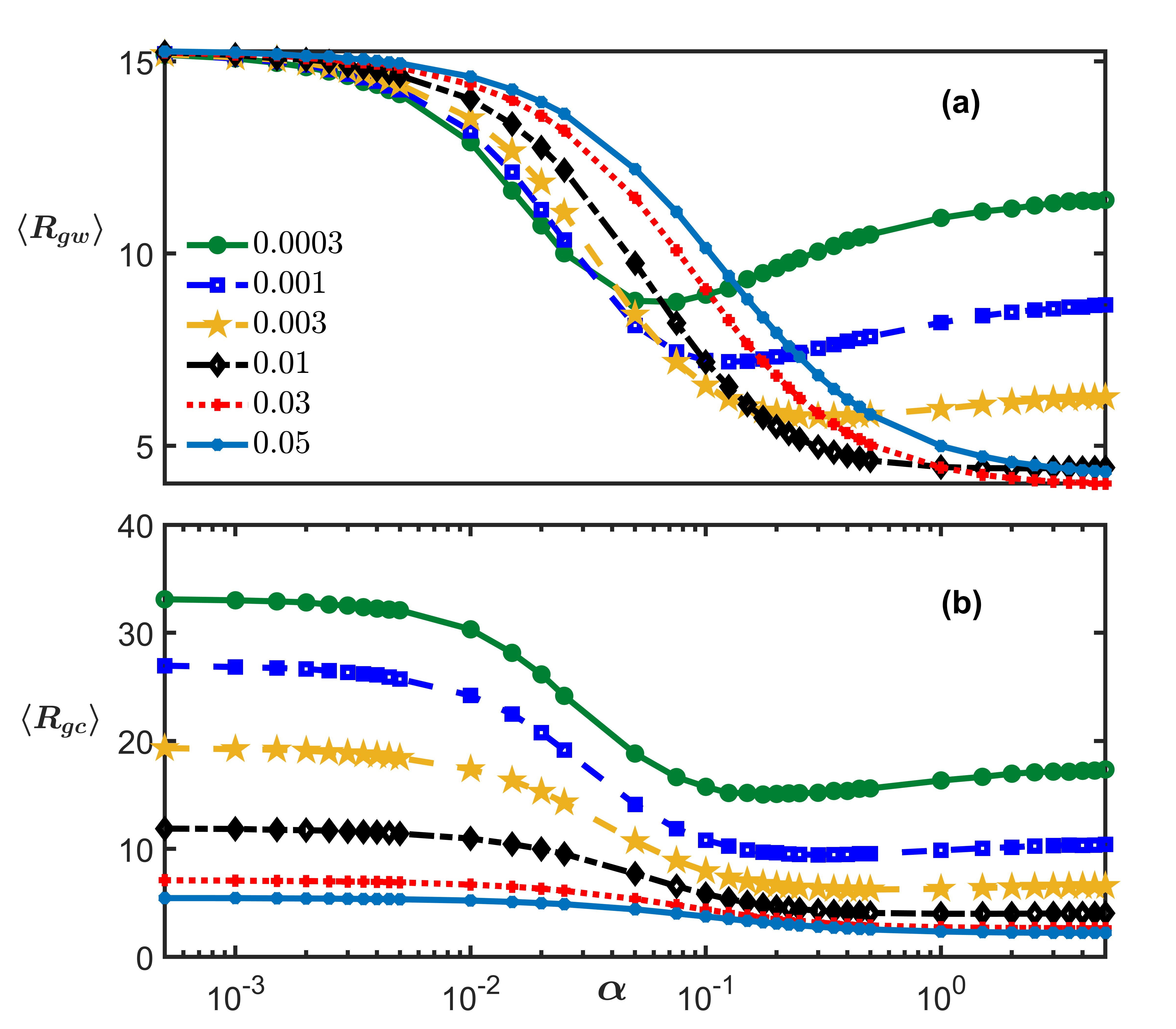}
\caption{Semi-log plot of the averaged $R_g$ of (a) the walker  trajectory $\la R_{gw} \ra$ and (b) the chemical trail $\la R_{gc} \ra$ as a function of the deposition rate $\alpha$ for six different decay rates.}
\label{fig:crossover}    
\end{figure}

To describe typical properties of the resultant trajectories, unless specified otherwise, we study $N=512$ steps of $10^5$ independent walks fixing $\lambda=10$. We first analyze the trajectories by varying the chemical deposition strength $\alpha$. The decay rate $\beta$ is a property of the environment. We study phase diagrams identifying {\em coil} and {\em globule} states of the trajectories that we describe in detail in the following section. In Fig.~\ref{fig:config}(b-e), we have plotted typical trajectories of the walker and chemical trails in these states.

\begin{figure*}[!t]
    \centering
    \includegraphics[width=0.85\linewidth ]{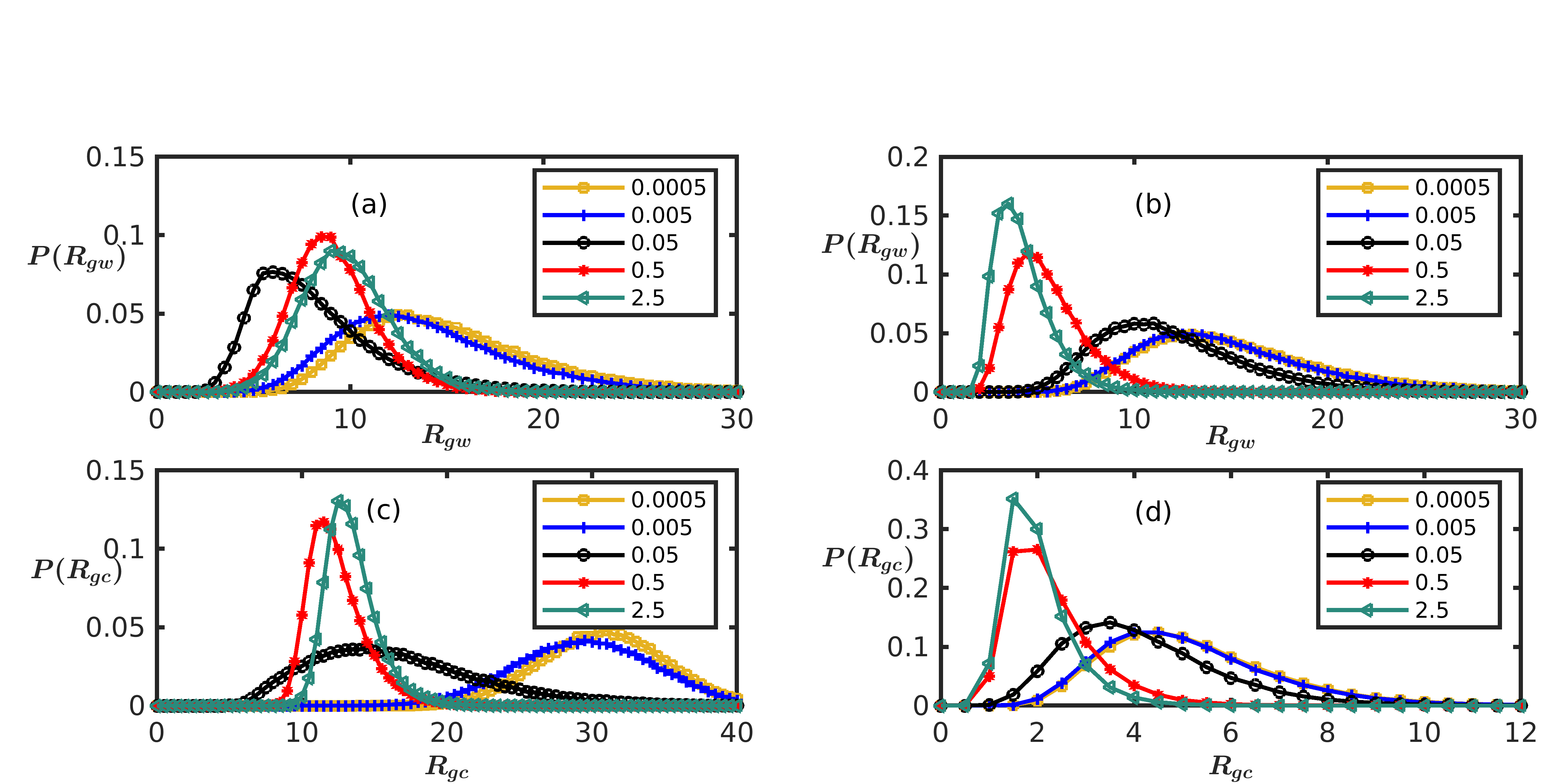}
    \caption{Probability distributions of the radius of gyration of the trajectories of the walker $P(R_{gw})$ and the corresponding chemical trail $P(R_{gc})$ for five different $\alpha$ values are plotted for (a-c) $\beta = 0.0005$ and  (b-d) $\beta = 0.005$.}
    \label{fig:prob}
\end{figure*}

\section{Results}
The walker trajectories and the deposited chemical profile $c_i(t)$ evolve together. Depending on the deposition and evaporation rates of the chemical cue we find either extended or highly localized trajectories. The transition between such conformations of walker paths are reminiscent of the coil-globule transition in polymers. In Flory-Huggins theory polymers undergo a coil to globule transition with decreased solubility of the medium that can lead to effective attraction between polymer segments~\cite{de1979scaling, deGennes1975}. Explicit binding between polymer segments also leads to similar transitions~\cite{LeTreut2016, Kumar2019}. In the current scenario, the chemical cue leads to effective attraction between earlier and later segments of the walker trajectory. 

\begin{figure}[!t]
    \centering
    \includegraphics[width=0.95\linewidth]{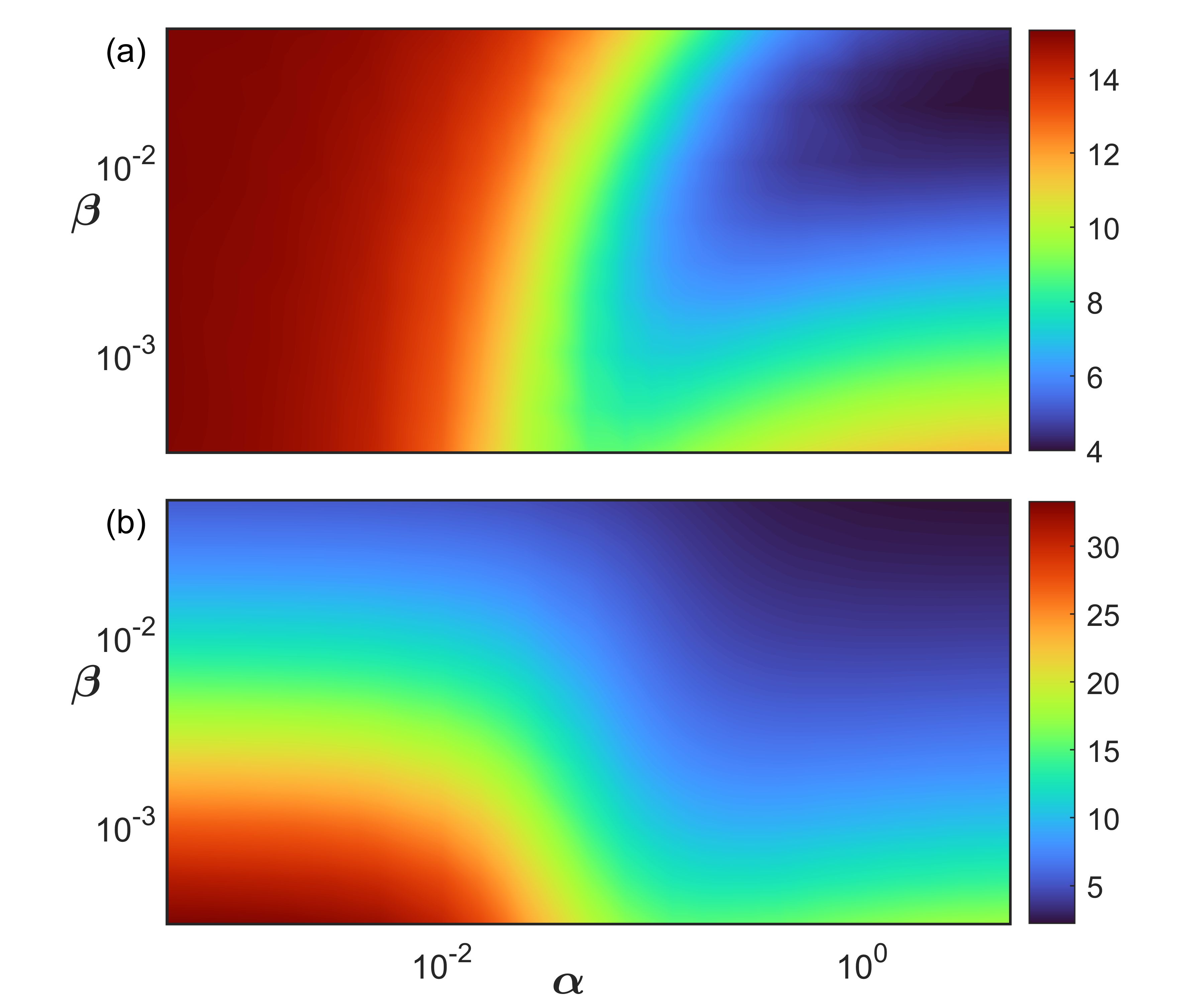}
    \caption{Phase diagrams calculated from the probability distributions of the radius of gyration for (a) the walker trajectory and (b) the chemical trail for a range of values of $\alpha$ and $\beta$. Color code indicates the value of the average $R_g$ with red denoting high $R_g$ and blue denoting low $R_g$.}
    \label{fig:phaseRg}
\end{figure}

The extensions of finite time trajectories sharply change their characteristic from open to compact conformations with change in deposition and evaporation rates of chemical cue.  
In the limit of $\a \ll \be$, the trajectories are extended and random-walk like. The chemical trail follows the walker trajectory, and evaporates quickly~(Fig.~\ref{fig:config}(b,c)). On increasing $\a$, the walker gets locally trapped in the chemical trail forming more compact configurations~(Fig.~\ref{fig:config}(d,e)). With $\a$ the trajectories undergo a {\em crossover} from extended coils to  compact globules. In the following, we study such crossovers with changing $\a,\be$ and obtain phase diagrams for the walker and chemical trail extensions using the radius of gyrations.   

\vskip 0.3cm
\noindent 
{\em Radius of gyration}. 
The extension of walker trajectories and chemical trails can be measured in terms of the radius of gyration $R_g$ which is defined as the averaged square distance between the various segments $\rv_i$ of such trajectories and that of the centre of mass $\rv_{cm}= (1/N)\sum_{i=1}^N\rv_i$. We use 
\begin{equation}
 R_g^2 = \frac{1}{N}\sum_{i=1}^N ({\rv_i} - \rv_{cm})^2.  
 \label{5.1}
\end{equation}

In Fig.~\ref{fig:crossover}(a) we have plotted the radius of gyration of the walker $\la R_{gw}\ra$ averaged over $10^5$ trajectories, as we vary the rate $\alpha$ for fixed values of $\beta$. 
At low $\alpha$, $\la R_{gw} \ra$ is high as expected for simple random walk. With increase in deposition rate $\alpha$, $\la R_{gw} \ra$ undergoes a non-monotonic change. It reduces to a low value corresponding to a locally trapped  globule-like state. Finally, for very large $\alpha$, it may again increase as the neighborhood of the walker gets increasingly homogeneous with respect to the presence of chemical cues. We see a similar non-monotonic variation of the radius of gyration of the chemical trails $\la R_{gc} \ra$ as well~(Fig.~\ref{fig:crossover}(b)). Thus, $R_g$ serves as an excellent quantifier of the structural crossover of the trajectories and the chemical trails and is the correct order parameter to describe the system behavior.

\vskip 0.3cm
\noindent 
{\em Probability distributions of $R_g$}.
To characterize the nature of the crossover, we consider the probability distribution of $R_g$ for the particle trajectory ($P(R_{gw})$) and the chemical trail ($P(R_{gc})$) with varying deposition rate $\alpha$ at fixed values of evaporation rate $\beta$.

In Fig.~\ref{fig:prob}(a and c), we plot $P(R_{gw})$ and $P(R_{gc})$ for a very low decay rate $\beta$. For low values of $\alpha$, the distribution of the trajectory is flatter and peaks at higher $R_g$ indicative of the coiled state. With increasing $\alpha$, the distribution becomes sharper and the peak shifts to lower values of $R_g$ indicative of the globule state. We also note the non monotonicity in the shifting of the peak of the distribution with increasing $\alpha$, reiterating the re-entrant feature discussed before. While the distribution of the chemical trail follows the same general features as that of the trajectory,  although the nonmonotonic feature is less pronounced. The absence of bimodality, which is a signature of phase coexistence, suggests that the coil-globule transition is continuous in this system. 

In Fig.~\ref{fig:prob}(b and d), we plot $P(R_{gw})$ and $P(R_{gc})$ for a relatively higher decay rate. Again the characteristic features of the distributions remain the same. The difference is that unlike in the case of low $\beta$, peak of the trajectory distribution $P(R_{gw})$ monotonically shifts to lower values with increasing $\alpha$.

\vskip 0.3cm
\noindent 
{\em Phase diagrams.} Having discussed the details of the structural transition of the trajectories of the walker in the self-generated chemical field, we now plot the phase diagrams of the walker and the chemical corresponding to the range of values of $\alpha$ and $\beta$. Fig.~\ref{fig:phaseRg}(a) gives the phase diagram of the walker trajectory. The color code indicates increasing values of the order parameter $\la R_g \ra$ from blue to red.  Evidently at higher values of $\beta$, the walker trajectories goes from a coiled state to a globule state with increasing deposition rate. At very low values of the decay rate however, we see the reappearance of a coiled state at very high $\alpha$. This signifies a re-entrance behavior. 

Fig.~\ref{fig:phaseRg}(b) gives the corresponding phase diagram of the chemical trail. The chemical trail does not show any re-entrance feature at higher $\alpha$. However, unlike the walker trajectory, at a low value of $\beta$, the chemical trail shows a transition from a coiled state to a globule state with increasing $\beta$. At high $\beta$, the chemical trail remains in a globule state for all values of $\alpha$.

To understand the re-entrant feature of the coil-globule transition, we note that the probability of the walker to move to a site saturates beyond a certain amount of chemical concentration in that site. Therefore for low decay rates $\beta$, as the deposition rate $\alpha$ is increased, the sites visited by the walker reaches this critical chemical concentration. If we further increase the deposition rate, the environment seen by the walker is uniform at this critical concentration. Therefore, the probabilities of moving to neighbouring sites is no longer biased and the walker can diffuse giving rise to a more extended walk similar to a coiled state.

\noindent
\section{Dynamics}
We have so far analysed the structural properties of the trajectories of the walker influenced by a chemical it secretes and gets attracted to. This showed a coil-globule transition in the extension of the trajectories  and a re-entrant behavior at larger $\alpha$ for low decay rates. In this section we analyze the dynamics of the system by considering the mean squared displacement $\la\Delta r^2(t)\ra$ of the walker. 

\begin{figure}[t] 
    \centering
    \includegraphics[width=0.8\linewidth]{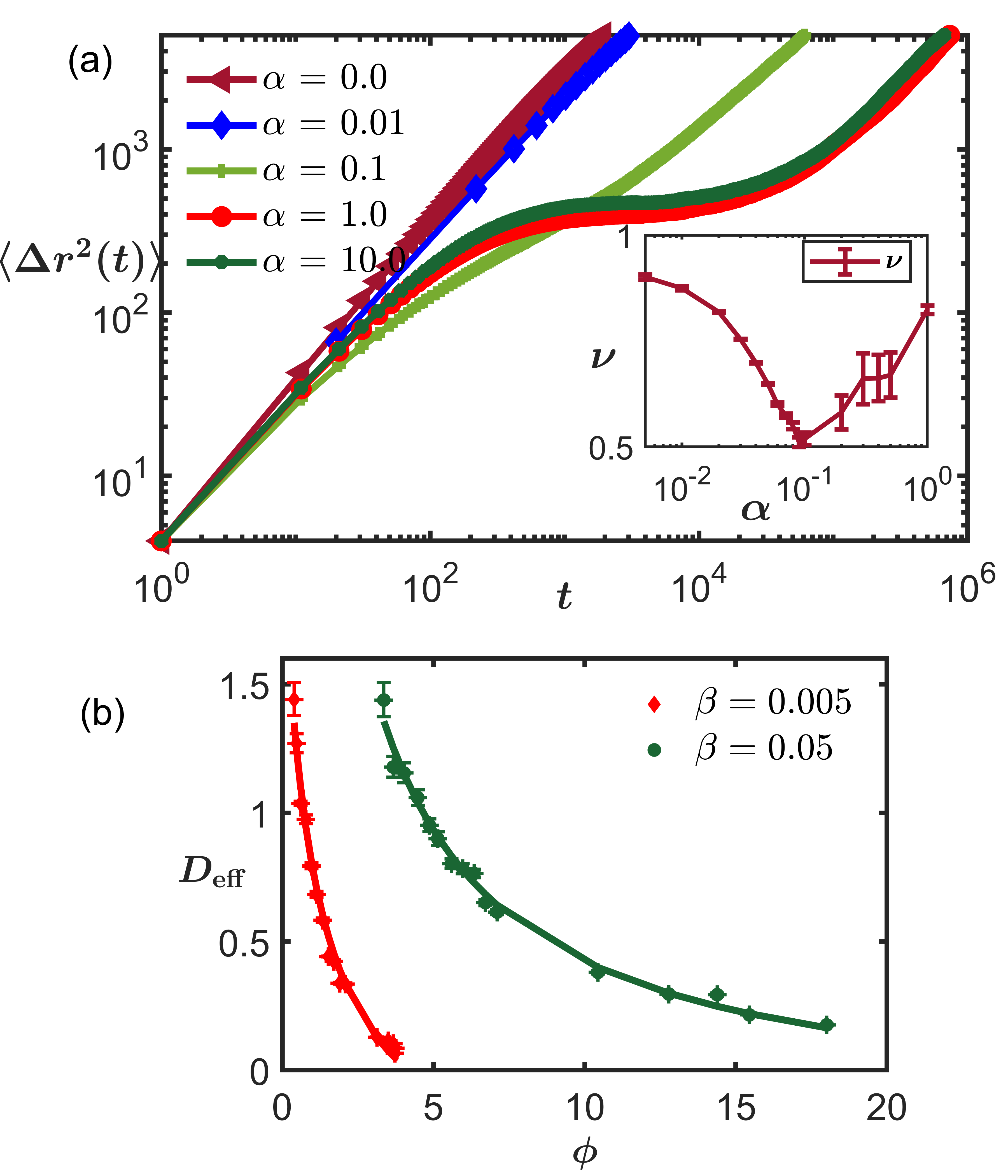}
    \caption{(a) Log-log plot of $\la\Delta r^2(t)\ra$ vs t for fixed $\beta = 0.0005$. Inset shows the plot of the exponent, $\nu$ from the relation $\la\Delta r^2(t)\ra \sim t^{\nu}$ as a function of $\alpha$ for low $t$.
    (b) $D_{\rm eff}$ obtained from simulations (points) and their fitting to Eq.(\ref{eq_D_phi}) (lines) are  plotted as a function of $\phi$ for two values of $\beta$. $\phi = k/R_{gc}^2$ is substituted in Eq.~\ref{eq_D_phi} to obtain $D_{\rm eff}$. 
    Eq.(\ref{eq_D_phi}) fits well with simulation results with fitting parameters $k \approx 100$ and $D=2.07, \l' = 2.564$ for $\beta = 0.005$ and  $D = 6.553$, $\l'$ = 6.744 for $\beta = 0.05$, respectively. 
    }
    \label{fig:msd}
\end{figure}

In Fig.~\ref{fig:msd}, we plot $\la\Delta r^2(t)\ra$ for various $\alpha$ at a fixed value of the decay rate. Even at relatively small $t$ ($t \lesssim 10^3$), $\la\Delta r^2(t)\ra$ shows distinctly different behavior as $\alpha$ is varied. To quantify the behavior, we fitted the data using $\la\Delta r^2(t)\ra \sim t^{\nu}$ for the exponent $\nu$. In Fig.~\ref{fig:msd}(a)(inset), we show $\nu$ as a function of $\alpha$. For very low $\alpha$, we observe a diffusive behavior with $\nu \approx 1$. As $\alpha$ is increased, the walk becomes sub-diffusive with $\nu < 1$. With further increase in $\alpha$, the exponent $\nu$ starts to increase again towards a diffusive scaling. This non-monotonic variation of $\nu$ with $\alpha$ correlates with the coil-globule-coil transition observed at low $\beta$. As the walker trajectory changes from a coiled to globule state, the mean squared displacement goes from a diffusive to a sub-diffusive behavior. At higher $\alpha$, as trajectories move back towards a coiled state, $\la\Delta r^2(t)\ra$ shows a more diffusive behavior.

At intermediate times and higher $\alpha$, $\la\Delta r^2(t)\ra$ saturates indicating that the walker gets trapped. At longer $t$, the dynamics become diffusive for all values of $\alpha$ to show a scaling $\la\Delta r^2(t)\ra \sim D_{\text{eff}}t$, with the effective diffusivity $D_{\text{eff}}$. The effective diffusivity includes the modification due to the presence of the autonomous chemical field which generates a self-trapping reducing the diffusion. Note that the slope of the curve in Fig.~\ref{fig:msd} (a) vanishes at intermediate times for enabling parameter values $\alpha$ leading to a chemical capture. This intermediate capture by the local chemical concentration $\phi$ is released at later times as more and more regions reach equivalent amount of chemical allowing the MSD to move back to diffusive track. The variation of asymptotic $D_{\rm eff}$ with $\phi$ is discussed in detail in the following section.

\noindent
\section{Mean field description}

Here we develop the evolution of a density field $\r(\rv,t)$ of the particle dependent on the associated chemical field $\phi(\rv,t)$. The particle motion is described by the Langevin equation 
\bea
\dot\rv = \chi \nabla \phi + \sqrt{2D}\eta.
\label{lange1}
\eea
Before proceeding with the rest of the analysis, let us first build a direct relation between the mobility $\chi$ and the concentration dependent hopping parameter $\l$ used in the numerical simulations. Discretizing the drift current corresponding to Eq.(\ref{lange1}) over the $i$-th bond, we get $j^d_i = \r_i \, \f{\chi}{2a} [\phi_{i+1} - \phi_{i-1}]$. In this expression we can identify a velocity operator $v_i = \f{\chi}{2a} [\phi_{i+1} - \phi_{i-1}]$. In the simulations, the hopping rate $p_i$ can be utilized to obtain a similar velocity operator $2a [p_{i+1} - p_{i-1}] = \f{2 a \l }{{\cal Z}(1+c_i)}(c_{i+1}-c_{i-1})$, where ${\cal Z}$ is the $\l$-dependent normalization used in Eq.(\ref{eqn2}). Identifying the chemical concentration $c_i=\phi_i$, a comparison of the two expressions for velocity operator gives 
$
\chi \approx \f{\l}{{\cal Z}(1+c_i)} (2a)^2.
$
This analysis shows that the effective mobility $\chi$ is not a constant, rather it depends on the chemical concentration. In the continuum notation, expressing the dimensional constant $\l' = 4a^2 \l/{\cal Z}$ quantifying the displacement bias towards chemical concentration, we get 
\bea
\chi = \l'/(1+\phi).
\eea
On the other hand, the hopping probability $p_i=1/6$ is constant in both the limits of vanishing and high concentrations $c_i$ leading to simple diffusion. As a result the diffusivity $D$ is treated here to be  independent of chemical concentration.  

The particle density corresponding to the Langevin equation (Eq. \ref{lange1}) evolves as $\p_t \r(\rv,t) = - \nabla \cdot \jv$, where the conserved current $\jv = -D \nabla \r + \chi \r \nabla \phi$. 
This leads to the following driven diffusion equation~\cite{keller1970initiation,mahdisoltani2021}
\bea
\p_t \r = \nabla \cdot \left( D \nabla \r - \chi \r \nabla \phi \right)
\label{eqro}
\eea
The chemical field deposited by the agent evolves as 
\bea
\p_t \phi (\rv,t) = \a \r - \be \phi.
\label{eqphi}
\eea
This is a non-conserved dynamics and relaxes faster than the conserved dynamics of $\r(\rv,t)$. Thus one can use the adiabatic approximation $\phi = (\a/\be) \r$ to obtain $\p_t \r(\rv,t) = - \nabla \cdot \jv$ with $\jv = -D_{\rm eff}(\r) \nabla \r$ where 
\begin{equation}
D_{\rm eff}(\r) = D- \chi \r\frac{\alpha }{\beta} 
= D - \frac{\alpha }{\beta} \f{\l' \r}{1+(\a/\be)\r}
\label{eqn_Deff}
 \end{equation}

$D_{\rm eff}$ is the effective diffusion coefficient that depends on the concentration $\rho$.  
 Within the adiabatic approximation, the above expression can be rewritten as 
 \bea
 D_{\rm eff} = D-\l' \phi/(1+\phi). 
 \label{eq_D_phi}
 \eea
 At small concentration of chemical cue the effective diffusivity decreases with the chemical concentration linearly to saturate for large concentrations. This non-linearity may stabilize $D_{\rm eff}$ to a positive value if $\l'<D$. A possibility of diffusion edge can appear if $D_{\rm eff}$ approaches zero asymptotically~\cite{golestanian2019bose,meng2021magnetic}. Otherwise, a $\l'>D$ can potentially lead to an instability towards localization. 
 The form of $D_{\rm eff}$ in Eq.(\ref{eqn_Deff}) is compared with the simulation results for $D_{\rm eff} $ in Fig. \ref{fig:msd} (b). In this figure, we used an estimate of the chemical concentration $\phi$ from the extension of chemical trajectories $R_{gc}$, using $\phi=k/R_{gc}^2$ where we treat $k$ as a fitting constant. 
 
 The agreement of simulation results for $D_{\rm eff} $ with Eq.(\ref{eq_D_phi}) shows that our mean-field theory explains the behaviour of the self-chemotactic walker in the simulation with good accuracy.  
 The expected values of D and $\l'$ are of the same order of magnitude as obtained from the fit. The estimates in the two examples shown in Fig.\ref{fig:msd}(b) give $D<\l'$ indicating a potential instability and not a diffusion edge. We present a linear stability analysis in the following section to analyze such a possibility.

\subsection{Linear stability analysis}
We perform a linear stability analysis of the two component system around a spatially homogeneous steady state with $\r=\r_0$, $\phi=\phi_0$. Linearizing Eq.(\ref{eqro}) and Eq.(\ref{eqphi}) around this steady state using $\r=\r_0 +\d \r$ and $\phi=\phi_0+\d \phi$ gives, 
\begin{equation}
    \begin{aligned}
     \partial_t \delta \rho &= D\nabla^2 \delta \rho - \chi\r_0\nabla^2 \d \phi \\
    \partial_t \delta \phi &=\a\d\r - \beta \delta\phi
\end{aligned}
\label{eq.14}
\end{equation}
Performing a Fourier transform of these equations gives the evolution of the modes in the form of the following matrix equation, $\partial_t(\delta \r_q, \delta \phi_q) = {\mathcal{M}}(\delta \r_q, \delta \phi_q)$ where ${\mathcal M}_{11} = -q^2D,{\mathcal M}_{12} = \r_0\chi q^2, {\mathcal M}_{21} = \a$ and ${\mathcal M}_{22} = -\beta$.
The eigen values of $\mathcal{M}$ are given by
\bea
\lambda(q^2) = \frac{1}{2}\{\text{Tr} \mathcal{M} \pm \sqrt{\text{Tr} \mathcal{M} - 4 \text{det}\mathcal{M}} \}.
\eea
where $\text{Tr}$ and $\text{det}$ denotes the trace and determinant of the matrix ${\mathcal{M}}$. Since, Tr$\mathcal{M} = -q^2D - \beta$ $< 0$, the only way to have instability is det$\mathcal{M}$ $< 0$. Thus, the condition for instability is given by
 \bea
     q^2(D\beta - \a \r_0 \chi) < 0.
 \eea
 Note from Eq.(\ref{eqn_Deff}) that this criterion is equivalent to $D_{\rm eff} < 0$.  Fig.~\ref{fig:msd} (b) showed that parameter values used in the simulations give $D<\l'$, i.e., $D_{\rm eff} < 0$ at large $\phi$ potentially allowing a linear instability towards localization. In other words, this requires a large coupling strength for the walker to follow its own chemical trail, $\chi > D/\phi$. Such a possibility of localization instability is similar to the prediction in Ref.~\cite{Tsori2004}. However, as Fig.~\ref{fig:msd} (b) shows, the self-generated chemical concentration $\phi$ does not increase to such large values that could potentially trap the walker asymptotically. Rather, the value remains restricted to a regime where $D_{\rm eff}>0$, although with small diffusion constant, keeping the walker diffusive at late times. Thus, although our theory allows for a localization via chemical secretion mediated self-interaction, our numerical simulations show that the walker does not get into that regime. This is  in agreement with the findings of Ref.~\cite{Grima2005, Grima2006}. 
  
  The change in the asymptotic diffusivity due to the coupling with chemical field in auto-chemotaxis has been discussed earlier in the context of persistent motion of active particles~\cite{Taktikos2011, Kranz2016, Kranz2019}. However, in that context the physical processes involved were different. The change in orientation in heading direction due to the chemical coupling led to the reduction of effective active diffusion.  In contrast, within our model, the chemical coupling generates a bias for particle accumulation towards higher density countering the diffusion current. This leads to a reduction of effective diffusivity with particle concentration. In models without chemical evaporation the walker can get localized~\cite{Kranz2016, Kranz2019, Tsori2004}, a possibility that disappears once evaporation is introduced~\cite{Taktikos2011}.

\section{Discussion and outlook}

  In this paper, we have studied the behaviour of a single random walker secreting a chemical which acts as local attractant to the walker. We performed direct numerical simulations to study the resultant pattern formation. Using this simple example we have developed a description of the resultant pattern in terms of the radius of gyration measuring the extent of the particle motion and chemical produced. There are two parameters that control the chemical environment of the walker, one is the rate at which chemical is 
  deposited ($\alpha$) and the rate of chemical evaporation ($\beta$) due to external factors.

  From numerical simulations, we observed a coil-globule transition in the trajectory of the walker and the chemical trail. The probability distribution of the $R_g$ remained unimodal across parameter values. The absence of multimodality signifies a continuous crossover from the coil to globule state. We developed a mean-field description of the coupled evolution of particle density and the concentration of chemical field. This led to an effective diffusivity that decreases with the concentration of chemical deposited. Our direct  numerical simulations showed good agreement with this analytic prediction. Further, our theory and simulation showed signatures of a linear instability towards increasing localization of the walker. A full stochastic field theory corresponding to our model will involve density-dependent multiplicative noise and remains an interesting direction of study.

  The model we considered is closely related to ants that perform persistent as well as directed motion. While moving, they release pheromone and respond to the pheromone profile in the environment. Current effort in our group is extending our model to persistent walkers, and studies of multiple walkers that can interact via the chemical profile as well as via local volume exclusion. 
  
  \section{Acknowledgments}
 S.S.K. thanks IOP, Bhubaneswar for hospitality during a visit that helped in initiating this project. S.S.K. and A.C. acknowledge the use of the computing facility at IISER Mohali. D.C. thanks SERB, India for financial support through Grant No. MTR/2019/000750 and International Centre for Theoretical Sciences (ICTS) for an associateship.
  
  \bibliographystyle{prsty.bst}

\begin{thebibliography}{10}

\bibitem{Moussaid2009}
M. Moussaid, S. Garnier, G. Theraulaz, and D. Helbing, Topics in Cognitive
  Science {\bf 1},    (2009).

\bibitem{Olsen2019}
M. Olsen, Stigmergy for Biological Spatial Modeling, 2019.

\bibitem{camazine2020}
S. Camazine, J. Deneubourg, N. Franks, J. Sneyd, G. Theraula, and E. Bonabeau,
  {\em Self-Organization in Biological Systems}, {\em Princeton Studies in
  Complexity} (Princeton University Press, ADDRESS, 2020).

\bibitem{Mukherjee2019}
S. Mukherjee and B.~L. Bassler, Nature Reviews Microbiology {\bf 17},  371
  (2019).

\bibitem{John2009}
A. John, A. Schadschneider, D. Chowdhury, and K. Nishinari, Phys. Rev. Lett.
  {\bf 102},  108001  (2009).

\bibitem{Chaudhuri2015}
D. Chaudhuri and A. Nagar, Phys. Rev. E {\bf 91},  012706  (2015).

\bibitem{Helbing1997}
D. Helbing, F. Schweitzer, J. Keltsch, and P. Molnár, Physical Review E {\bf
  56},    (1997).

\bibitem{Vicsek1994nature}
E. Ben-Jacob, O. Schochet, A. Tenenbaum, I. Cohen, A. Czirók, and T. Vicsek,
  Nature {\bf 368},    (1994).

\bibitem{Chaudhuri2011}
D. Chaudhuri, P. Borowski, and M. Zapotocky, Physical Review E - Statistical,
  Nonlinear, and Soft Matter Physics {\bf 84},    (2011).

\bibitem{Robles2018}
A.~H. Robles and L. Giuggioli, Physical Review E {\bf 98},    (2018).

\bibitem{Feinerman2017}
O. Feinerman and A. Korman, Journal of Experimental Biology {\bf 220},
  (2017).

\bibitem{Khuong2016}
A. Khuong, J. Gautrais, A. Perna, C. Sbaï, M. Combe, P. Kuntz, C. Jost, and G.
  Theraulaz, Proceedings of the National Academy of Sciences {\bf 113},
  (2016).

\bibitem{Gloag2016}
E.~S. Gloag, L. Turnbull, M.~A. Javed, H. Wang, M.~L. Gee, S.~A. Wade, and
  C.~B. Whitchurch, Scientific Reports {\bf 6},    (2016).

\bibitem{king2009chemotaxis}
J.~S. King and R.~H. Insall, Trends in cell biology {\bf 19},  523  (2009).

\bibitem{mittal2003motility}
N. Mittal, E.~O. Budrene, M.~P. Brenner, and A. Van~Oudenaarden, Proceedings of
  the National Academy of Sciences {\bf 100},  13259  (2003).

\bibitem{jin2017chemotaxis}
C. Jin, C. Kr{\"u}ger, and C.~C. Maass, Proceedings of the National Academy of
  Sciences {\bf 114},  5089  (2017).

\bibitem{Rutherford2012}
S.~T. Rutherford and B.~L. Bassler, Cold Spring Harbor Perspectives in Medicine
  {\bf 2},  a012427  (2012).

\bibitem{Laganenka2016}
L. Laganenka, R. Colin, and V. Sourjik, Nature Communications {\bf 7},
  (2016).

\bibitem{helbing1997modelling}
D. Helbing, J. Keltsch, and P. Molnar, Nature {\bf 388},  47  (1997).

\bibitem{Bechinger2016}
C. Bechinger, R.~D. Leonardo, H. Löwen, C. Reichhardt, G. Volpe, and G. Volpe,
  Reviews of Modern Physics {\bf 88},  045006  (2016).

\bibitem{Marchetti2013}
M.~C. Marchetti, J.~F. Joanny, S. Ramaswamy, T.~B. Liverpool, J. Prost, M. Rao,
  and R.~A. Simha, Rev. Mod. Phys. {\bf 85},  1143  (2013).

\bibitem{Malakar2018}
K. Malakar, V. Jemseena, A. Kundu, K. {Vijay Kumar}, S. Sabhapandit, S.~N.
  Majumdar, S. Redner, and A. Dhar, J. Stat. Mech. Theory Exp. {\bf 2018},
  043215  (2018).

\bibitem{Dolai2020}
P. Dolai, A. Das, A. Kundu, C. Dasgupta, A. Dhar, and K.~V. Kumar, Soft Matter
  {\bf 16},  7077  (2020).

\bibitem{Santra2021}
I. Santra, U. Basu, and S. Sabhapandit, Phys. Rev. E {\bf 104},  L012601
  (2021).

\bibitem{Shee2020}
A. Shee, A. Dhar, and D. Chaudhuri, Soft Matter {\bf 16},  4776  (2020).

\bibitem{Chaudhuri2020}
D. Chaudhuri and A. Dhar, J. Stat. Mech. Theory Exp. {\bf 2021},  013207
  (2021).

\bibitem{Shee2021b}
A. Shee and D. Chaudhuri, J. Stat. Mech. Theory Exp. {\bf 2022},  013201
  (2022).

\bibitem{Shee2022a}
A. Shee and D. Chaudhuri, Phys. Rev. E {\bf 105},  054148  (2022).

\bibitem{Lam1995}
L. Lam, Chaos, Solitons \& Fractals {\bf 6},    (1995).

\bibitem{Castillo1995}
V. Castillo, M. Veinott, and L. Lam, Chaos, Solitons \& Fractals {\bf 6},
  (1995).

\bibitem{Schweitzer2007}
F. Schweitzer, {\em Browning Agents and Active Particles} (Springer Berlin
  Heidelberg, ADDRESS, 2007).

\bibitem{rodriguez2001fractal}
I. Rodriguez-Iturbe and A. Rinaldo, {\em Fractal river basins: chance and
  self-organization} (Cambridge University Press, ADDRESS, 2001).

\bibitem{Chaudhuri2009}
D. Chaudhuri, P. Borowski, P.~K. Mohanty, and M. Zapotocky, EPL (Europhysics
  Lett. {\bf 87},  20003  (2009).

\bibitem{Chaudhuri2011e}
D. Chaudhuri, P. Borowski, and M. Zapotocky, Phys. Rev. E {\bf 84},  021908
  (2011).

\bibitem{Taktikos2011}
J. Taktikos, V. Zaburdaev, and H. Stark, Physical Review E {\bf 84},  041924
  (2011).

\bibitem{Kranz2016}
W.~T. Kranz, A. Gelimson, K. Zhao, G.~C. Wong, and R. Golestanian, Physical
  Review Letters {\bf 117},  038101  (2016).

\bibitem{Kranz2019}
W.~T. Kranz and R. Golestanian, The Journal of Chemical Physics {\bf 150},
  214111  (2019).

\bibitem{Grima2005}
R. Grima, Physical Review Letters {\bf 95},  128103  (2005).

\bibitem{Grima2006}
R. Grima, Physical Review E {\bf 74},  011125  (2006).

\bibitem{Sengupta2009}
A. Sengupta, S. van Teeffelen, and H. Löwen, Physical Review E {\bf 80},
  031122  (2009).

\bibitem{Tsori2004}
Y. Tsori and P.-G. de~Gennes, Europhysics Letters (EPL) {\bf 66},  599  (2004).

\bibitem{saha2014clusters}
S. Saha, R. Golestanian, and S. Ramaswamy, Physical Review E {\bf 89},  062316
  (2014).

\bibitem{marsden2014chemotactic}
E.~J. Marsden, C. Valeriani, I. Sullivan, M. Cates, and D. Marenduzzo, Soft
  Matter {\bf 10},  157  (2014).

\bibitem{hokmabad2022chemotactic}
B.~V. Hokmabad, J. Agudo-Canalejo, S. Saha, R. Golestanian, and C.~C. Maass,
  Proceedings of the National Academy of Sciences {\bf 119},  e2122269119
  (2022).

\bibitem{de1979scaling}
P.-G. De~Gennes and P.-G. Gennes, {\em Scaling concepts in polymer physics}
  (Cornell university press, ADDRESS, 1979).

\bibitem{deGennes1975}
P.~G. de~Gennes, J. Phys. Lettres {\bf 36},  55  (1975).

\bibitem{LeTreut2016}
G. {Le Treut}, F. K{\'{e}}p{\`{e}}s, and H. Orland, Biophys. J. {\bf 110},  51
  (2016).

\bibitem{Kumar2019}
A. Kumar and D. Chaudhuri, J. Phys. Condens. Matter {\bf 31},  354001  (2019).

\bibitem{keller1970initiation}
E.~F. Keller and L.~A. Segel, Journal of theoretical biology {\bf 26},  399
  (1970).

\bibitem{mahdisoltani2021}
S. Mahdisoltani, R.~B.~A. Zinati, C. Duclut, A. Gambassi, and R. Golestanian,
  Physical Review Research {\bf 3},  013100  (2021).

\bibitem{golestanian2019bose}
R. Golestanian, Physical Review E {\bf 100},  010601  (2019).

\bibitem{meng2021magnetic}
F. Meng, D. Matsunaga, B. Mahault, and R. Golestanian, Physical Review Letters
  {\bf 126},  078001  (2021).

\end{thebibliography}

\end{document}